\documentclass[useASM]{mn2e}
\input{psfig.sty}
\usepackage{graphicx}
\begin{document}

\title{Testing the Predictions of the Universal-Structured Jet Model of Gamma-Ray Bursts by Simulations
\footnote{send offprint request to: E. W. Liang (Email: ewliang@nju.edu.cn)}}

\author[E. W. Liang , X. F. Wu, and Z. G. Dai]
       {E. W. Liang$^{1,2,3}$,  X. F. Wu$^{1}$, and Z. G. Dai$^{1}$
 \\
        $^1$Astronomy Department, Nanjing University, Nanjing 210093, China; ewliang@nju.edu.cn\\
        $^2$Physics Department, Guangxi University, Nanning 530004, China\\
        $^3$National Astronomical Observatories/Yunnan Observatory, Chinese Academy of
Sciences, Kunming 650011, China}
%;\\
%        $^4$Physics Department, Jingdezhen Comprehensive College, Jingdezhen 333000, P.R. China\\
%        $^5$The Graduate School of the Chinese Academy of Sciences, Beijing, P. R. China\\
%        $^6$Physics Department, Xiangfan University, Xiangfan 441053, P.R. China}

\maketitle

\label{firstpage}

\begin{abstract}
Based on the standard energy reservoir of gamma-ray burst (GRB) jets and the relationship between
peak energy ($E_p$) of $\nu F_{\nu}$ spectrum and the equivalent-isotropic energy ($E_{\rm{iso}}$),
we test the GRB probability distributions as a function of viewing angle ($\theta$) and of viewing
angle together with redshift $z$, $P(\theta)$ and $P(\theta,z)$, predicted by the
universal-structured jet (USJ) model with simulations on observational bases for a detection
threshold of $S>4 \times 10^{-5}$ erg cm$^{-2}$, where $S$ is the total fluence in an energy band
$>20$ keV. We simulate a sample including $10^{6}$ GRBs, and then derive $P(\theta)$ and
$P(\theta,z)$. We show that both simulated and theoretical $P(\theta)$ and $P(\theta,z)$ are
consistent with each other. This consistency seems to be regarded as support for the USJ model. A
small difference between the theoretical and simulated results might be due to the observed $E_p$
distribution used in our simulations, which perhaps suffers a little of biases for $E_p<100$ keV.

\end{abstract}
\begin{keywords}
gamma rays: bursts---gamma rays: observations---ISM: jets and outflows---methods: statistical
\end{keywords}

\section{Introduction}               % Introduction goes below.
Significant progress on understanding the nature of gamma-ray bursts (GRBs) and their afterglows
has been made in the recent decade (for reviews see Fishman \& Meegan 1995; Piran 1999; van
Paradijs et al. 2000;  Cheng \& Lu 2001;  M\'{e}sz\'{a}ros 2002). However, this phenomenon is still
of a great mystery. One of the most difficult parts of solving this mystery is their central
engines, which are far away from emission regions and are hid behind our observations (e.g., Cheng
\& Lu 2001). It is believed that GRBs are produced by conical ejecta powered by the central
engines. Sharp breaks and/or quick decays in afterglow light curves are thought to be evidence for
collimated ejecta (e.g., Rhoads 1999; Sari, Piran, \& Halpern 1999; Dai \& Cheng 2001). Starting
with GRB 990123 such evidence has been rapidly growing up (e.g., Bloom, Frail, \& Kulkarni 2003 and
references therein). In a previous work we showed that the current observations of prompt gamma-ray
emissions and X-ray afterglow emissions satisfy the prediction of GRB jet (Liang 2004). Both the
dynamics and afterglow light curve behaviors of GRB jets were investigated numerically (Panaitescu,
M\'{e}sz\'{a}ros \& Rees 1998; Huang et al. 2000; Moderski, Sikora \& Bulik 2000; Granot \& Kumar
2003; Kumar and Granot 2003; Salmonson 2003), and some GRB afterglows of interest were also fitted
(Panaitescu \& Kumar 2001a, 2001b, 2002).

GRB jet structure models are currently under heavy debate. Two currently competing models are
universal-structured jet (USJ) model (e.g., M\'{e}sz\'{a}ros, Rees \& Wijers 1998; Dai \& Gou 2001;
Rossi, Lazzati, \& Rees 2002; Zhang \& M\'{e}sz\'{a}ros 2002a; Granot \& Kumar 2003; Kumar \&
Granot 2003; Panaitescu \& Kumar 2003; Wei \& Jin 2003) and uniform jet model (e.g., Rhoads 1999;
Frail et al. 2001). X-ray flashes (XRFs), which are thought to be the low-energy extension of
typical GRBs (Kippen et al. 2003; Lamb, Donaghy, \& Graziani 2003; Sakamoto et al. 2004), seem to
provide more jet structure signatures. If both GRBs and XRFs are the same phenomenon, any jet
structure model should present a unified description for GRBs and XRFs. Zhang et al. (2004a) showed
that the current GRB/XRF prompt emission/afterglow data can be described by a quasi-universal
Gaussian-like (or similar structure) structured jet with a typical opening angle of $\sim 6^\circ$
and with a standard jet energy of $\sim 10^{51}$ ergs. Based on {\em High Energy Transient Explorer
2} ({\em HETE-2}) observations, Lamb et al. (2003) proposed that the uniform jet model reasonably
describes the unified scheme of GRBs/XRFs. Very recently, the two-component jet model was advocated
by Berger et al. (2003a) based on the observations of GRB 030329, which has two different jet
breaks in an early optical afterglow light curve (0.55 days; Price et al. 2003) and in a late radio
light curve (9.8 days; Berger et al. 2003a). Millimeter observations further support the
two-component jet in this burst (Sheth et al. 2003). Simulations of the propagation and eruption of
relativistic jets in massive Wolf-Rayet stars by Zhang, Woosley, \& Heger (2004b) also showed
signatures of the two-component jet. In a previous work, we found a bimodal distribution of the
observed peak energy ($E_p$) of $\nu F_{\nu}$ spectra of GRBs/XRFs, and concluded that the
two-component jet model can explain this distribution (Liang \& Dai 2004). Our results suggest that
the two-component jet seems to be universal for GRBs/XRFs. Numerical calculations of such a model
were also presented by Huang et al. (2004).

The USJ model has a predictive power. This might be served as a test for this model. Perna, Sari,
\& Frail (2003) found that both the observational and theoretical probability distributions of
$\theta$, $P(\theta)$, are roughly consistent with each other. However, Nakar, Granot, \& Guetta
(2004) showed that the theoretically 2-dimensional probability distribution, $P(z,\theta)$, poorly
agrees with the observational one. We note that both Perna et al. (2003) and Nakar et al. (2004)
came to conclusions from the comparison of the USJ model's predictions with the current
observations. Whereas, the current sample of GRBs with $\theta$ available is very small (16 GRBs in
their works). Observational biases and sample incompleteness affect greatly this sample. It is
difficult to draw a robust conclusion in a statistical sense from the comparison of $P(\theta)$ or
$P(\theta, z)$ derived from this sample with that predicted by the USJ model. In addition, both
theoretical $P(\theta)$ and $P(\theta, z)$ are sensitive to instrument's threshold setting. In
Perna et al. (2003) and Nakar et al. (2004), the threshold is the same as CRGO/BATSE. However, most
of the bursts with $\theta$ available in the current GRB sample were not observed by CGRO/BATSE.
This difference also affects greatly the comparison results. We therefore make a further test by
simulations. Based on the standard energy reservoir of GRB jets (Frail et al. 2001; Bloom et al.
2003; Panaitescu \& Kumar 2001a; Piran et al. 2001; Berger, Kulkarni, \& Frail 2003b) and the
relationship between $E_p$ and $E_{\rm{iso}}$, the equivalent-isotropic energy (Amati et al. 2002;
Lloyd-Ronning \& Ramirez-Ruiz 2002; Atteia 2003; Sakamoto et al. 2004; Lamb et al. 2003; Liang, Dai
, \& Wu 2004; Yonetoku et al. 2004), we test the $P(\theta)$ and $P(\theta,z)$ predicted by the USJ
model with simulations on observational bases for a detection threshold of $S>4 \times 10^{-5}$ erg
cm$^{-2}$, where $S$ is the total fluence in an energy band $>20$ keV. We show that both
$P(\theta)$ and $P(\theta, z)$ derived from simulations and from the USJ model are consistent with
each other.

This paper is arranged as follows. The empirical model is presented in section 2. Our simulation
procedure is described in section 3. The results of the comparison of our simulation results with
that predicted by the USJ model are presented in sections 4. Discussion and conclusion are
presented in section 5. Throughout this work we adopt $H_0=65$ km s$^{-1}$ Mpc $^{-1}$,
$\Omega_{\rm{m}}=0.3$, and $\Omega_\Lambda=0.7$.

\section{Model}
Our simulation analysis model is based on the statistical results of a standard energy reservoir in
GRB jets and the relationship between $E_p$ and $E_{\rm{iso}}$. The energy release of a GRB jet in
the gamma-ray band is given by
\begin{equation}
\varepsilon_{52}=E_{\rm{iso,52}}(1-\cos\theta),
\end{equation}
where $E_{\rm{iso,52}}=E_{\rm{iso}}/10^{52} {\rm{ergs}}$. The relationship between $E_{\rm{iso}}$
and $E_{\rm{p}}$ is given by
\begin{equation}
E_{{\rm{iso,52}}}=k[E_{{\rm{p,2}}}(1+z)]^2,
\end{equation}
where $k$ is a coefficient, and $E_{{\rm{p,2}}}=E_{{\rm{p}}}/10^{2}{\rm{keV}}$. This empirical
relation was further investigated in standard synchrotron/inverse-Compton/synchro-Compton models
(Zhang \& M\'{e}sz\'{a}ros 2002b; Ramirez-Ruiz \& Lloyd-Ronning 2002). Sakamoto et al. (2004) and
Lamb et al. (2003) pointed out that {\em HETE-2} observations not only confirm this correlation,
but also extend it to XRFs. In a recent work, we also proved that this relationship holds within a
burst (Liang, Dai, \& Wu 2004). From Eqs. (1) and (2), we obtain
\begin{equation}
\theta={\rm{arccos}}\{1-\frac{\varepsilon_{52}}{k[E_{\rm{p,2}}(1+z)]^2}\}.
\end{equation}
Since the values of $k$ and $\varepsilon_{52}$ can be obtained by the current GRB sample, the
$\theta$ value of a burst can be estimated by Eq. (3), if its $E_p$ and $z$ are available. If both
distributions of $E_p$ and $z$ can be established by observations or theory, one can perform
simulations to derive a GRB sample at the least observational bias, and obtain $P(\theta)$ and
$P(\theta, z)$ from the simulated sample. Such a $E_p$ distribution for a given detection threshold
may be established by the BATSE observations. The redshift distribution is impossible to be
modelled by the presented GRB sample because of various observational biases, such as GRB
detection, localization, and redshift measurement (Bloom 2003). Please note that, in our analysis,
we focus on the comparison of the simulated $P(\theta)$ and $P(\theta, z)$  with that predicted by
the USJ model. If we use the same redshift distribution model in our simulations and in the USJ
model, the comparison results are not significantly affected by the redshift distribution (for a
detailed discussion in section 5). Thus, we perform simulations to derive $P(\theta)$ and
$P(\theta,z)$ and compare them to the predictions of the USJ model for a given detection threshold,
assuming the GRB rate as a function of redshift is proportional to the star formation rate. We
present the distributions of these parameters used in our simulations in the following.

The strong correlations shown in Eqs. (1) and (2) indicate that both $\varepsilon_{52}$ and $k$
have a standard candle feature. They should not be significantly affected by observational biases
and the sample selection effect. Their distributions can be modelled by the present GRB sample.
>From a sample of GRBs with $\theta$ available presented by Bloom et al. (2003), we find that the
$\varepsilon_{52}$ distribution  can be modelled by a log-normal one with a center at $-0.9$ and a
$\sigma_{\log \varepsilon_{52}}=0.3$, i.e.,
\begin{equation}
P(\log \varepsilon_{52})=P_{\varepsilon,0}e^{-2\frac{(\log \varepsilon_{52}+0.9)^2}{0.3^2}}
\end{equation}
where $P_{\varepsilon,0}$ is a normalized constant given by $\int^{\infty}_{-\infty}P(\log
\varepsilon_{52})d \log \varepsilon_{52}=1$. From a GRB sample presented by Amati et al. (2002), we
find that the $k$ distribution can be modelled by

\begin{equation}
P(k)=P_{\rm{k,0}}e^{-2\frac{(k-1)^2}{0.4^2}},
\end{equation}
where $P_{\rm{k,0}}$ is a normalized constant given by $\int^{\infty}_{0}P(k) d k=1$.

The observed $E_{\rm{p}}$ distribution is significantly affected by observational biases and sample
selection effects, especially when the completeness at low fluxes is considered. We modelled the
$E_p$ distribution for a given detection threshold from GRB observations. In a previous paper, we
studied the observed $E_{\rm{p}}$ distribution of GRBs and XRFs, combined with both {\em HETE-2}
and BATSE observations. We found that the observed $E_{\rm{p}}$ distribution for GRBs/XRFs is a
bimodal one with peaks of $\sim 30$ keV and $\sim 200$ keV (Liang \& Dai 2004). The $\sim 30$ keV
peak has a sharp cutoff at its lower energy side because of the limitation of the current
instrument thresholds. This bimodal distribution should be further examined by future observations.
In addition, for some extremely soft XRFs, they violate the model used in our analysis (Eq. 3;
e.g., Liang \& Dai 2004). We therefore cannot use this $E_p$ distribution in our analysis. It is
well known that the observed $E_p$ distribution of typical GRBs is narrowly clustered at 200-400
keV (e.g., Preece et al. 2000). The $E_p$ distribution of bright GRBs has been well established.
Thus, we consider only bright GRBs in our analysis. Preece et al. (2000) presented the observed
$E_{\rm{p}}$ distribution for a long, bright GRB sample selected by $S>4\times 10^{-5}$ erg cm
$^{-2}$ in an energy band $>20$ keV. We use this GRB sample to model the $E_p$ distribution, which
is

\begin{equation}
P(\log E_{p,2})=P_{E,0}e^{-2\frac{(\log E_{p,2}-0.38)^2}{0.45^2}},
\end{equation}
$P_{\rm{E,0}}$ is a normalized constant, which is given by $\int^{\infty}_{-\infty}P(\log E_{p,2})
d \log E_{p,2}=1$.

The GRB rate as a function of redshift is assumed to be proportional to the star formation rate
(e.g., Bromm \& Loeb 2002 ). The SF1 model from Porciani \& Madau (2001) is used in our analysis.
Thus, the redshift distribution is modelled by

\begin{equation}
P(z)=P_{\rm{z,0}}\frac{0.3e^{3.4z}}{e^{3.8z}+45},
\end{equation}
where $P_{\rm{z,0}}$ is a normalized constant. Since the largest redshift of GRBs is 4.5, we
restrict $z\leq 4.5$, and $P_{\rm{z,0}}$ is given by $\int^{4.5}_{0}P(z)dz=1$.

\section{Simulation Procedure}
Based on the empirical model and its parameter distributions presented above, we simulate a sample
including $10^{6}$ GRBs. Each simulated burst is characterized by a set of ($k$,
$\varepsilon_{52}$, $E_{\rm{p,2}}$, $z$), and it also satisfies our threshold setting. The
threshold setting corresponds to the sample selection criterion of the sample in Preece et al.
(2000), which is $S>4 \times 10^{-5}$ erg cm$^{-2}$. Our simulation procedure is described as
follows.

(1) Derive the accumulative probability distributions of the parameters from Eqs. (4)-(7), $Q(x)$,
where $x$ is one of these parameters. For saving calculation time, we use the discrete forms of
these distributions. By giving a bin size and truncating $x$ at a given value $a$, we calculate
$P(x_i)$ by Eqs. (4)-(7), where $x_{i}$ is the i$th$ bin of parameter $x$, and then derive
$Q(x_i)=\Sigma_{{i^{'}}=0,i}P(x_{i^{'}})$. The normalized constants in Eqs (4)-(7) are given by
$Q(x_i)|_{x_i=a}=1$.

(2) Simulate a GRB. We first generate a random number $m$ ($0<m\leq 1$), and obtain the value of
$x$ from the inverse function of $Q(x)=m$, i.e., $x=Q^{-1}(m)$. Please note that $Q(x)$ is in a
discrete form. The $x$ value for a given $m$ is restricted by $Q(x_{i})<m$ and $Q(x_{i+1})>m$.
Then, the $x$ value is given by $x=(x_{i+1}+x_{i})/2$. Repeating this step for each parameter, we
get a simulated GRB characterizing by a set of ($k$, $\varepsilon_{52}$, $E_{\rm{p,2}}$, $z$). If
this burst violates $\varepsilon_{52}/k[E_{\rm{p,2}}(1+z)]^2<1$, it does not satisfy Eq. (3) and is
ruled out without further consideration.

(3) Examine whether or not the simulated GRB satisfies our threshold setting. For a simulated GRB,
we calculate its flux in the observer frame by $F=E_{\rm{iso}}(1+z)/4\pi D_{L}^{2}(z)T$, where
$D_{L}(z)$ is the luminosity distance at $z$, $T$ is the ``effective" duration of the burst, and
$E_{\rm{iso}}$ is given by Eq. (2). The ``effective" duration in Perna et al. (2003) is 8 seconds.
In our analysis we take $T=12$ seconds (for a detail discussion in section 4). The limit of $F$ is
given by $F_{\rm{lim}}=S/T=3.3\times 10^{-6}$ erg cm$^{-2}$ s$^{-1}$. If $F>F_{\rm{lim}}$, the
burst is included into our mock GRB sample.

(4) Repeat steps (2) and (3) to obtain a mock sample of $10^{6}$ GRBs.

\section{Results}
We calculate the $\theta$ value for each mock GRB by Eq. (3). The relative probability of $\theta$
is shown in Figure 1 (the step line). For comparison, the results derived from the USJ model is
also plotted in Figure 1 (the straight line). Please note that, when we calculate the $\theta$
distribution by the USJ model, the ``effective" duration $T$ and the threshold setting are
different from that used in Perna et al. (2003). $T$ is an adjustable parameter, which is taken as
8 seconds in order to obtain a distribution that is agreement with the observational data in Perna
et al. (2003). From Figure 1, we can see that the simulated distribution is a log-normal one
centering at $\log \theta =-1.3$. This result is significantly different from the observational
result\footnote{Please note that the simulated $\theta$ distribution is dependent on the star
formation rate model and the detection threshold. It cannot be regarded as a {\em true} $\theta$
distribution. Any comparison between this distribution with observational results is
meaningless.}--The mean of $\theta$ for a sample of GRBs with $\theta$ available in Bloom et al.
(2003) is $\sim0.12$ rad, while our simulation result is clustered at $0.04$ rad. Hence, we do not
take the same $T$ value as that in Perna et al. (2003). We adjust its value to let both the
simulated $\theta$ distribution and that derived by the USJ model peak at the same position. We
obtain $T\sim 12$ seconds. For the threshold setting, we adopt $F_{\rm{lim}}=3.3\times 10^{-6}$, as
we mentioned in the above section. The maximum redshift ($z_{\rm{max}}$) up to which a burst with
apparent angle $\theta$ can satisfy the detection threshold is obtained from
\begin{equation}
F_{\rm{lim}}=\frac{E_{\rm{j}}}{4\pi D_{\rm{L}}^2(z_{\rm{max}})(1-\cos\theta)T}
\end{equation}
where $E_{j}=1.33 \times 10^{51}$ ergs (Bloom et al. 2003). Since the maximum redshift in current
GRB observations is 4.5, we set $z_{\rm{max}}=4.5$ when $z_{\rm{max}}>4.5$.

>From Figure 1, we find that the $\theta$ distribution of the mock GRB sample ranges in 0.001-0.25
rad, and the two distributions are well consistent when $\theta<0.10$ rad, while the simulated
$\theta$ distribution is lower than that predicted by the USJ model when $\theta>0.10$ rad. We
evaluate the consistency of these two distributions by a Kolmogorov-Smirnoff (K-S) test (Press et
al. 1997, p.617). The result of the K-S test is described by a statistic of $P_{\rm{K-S}}$: a small
value of $P_{\rm{K-S}}$ indicates a significant difference between two distributions
($P_{\rm{K-S}}=1$ indicates that two distributions are identical, and $P_{\rm{K-S}}<10^{-4}$
suggests that the consistency of two distributions should be rejected; e.g., Bloom 2003). The K-S
test shows that $P_{\rm{K-S}}=0.73$, indicating that the consistency of the two distributions is
acceptable.

Nakar et al. (2004) found that the 2-dimensional distribution of the probability from observations,
$P(\log \theta,\ z$), is poorly consistent with that predicted by the USJ model. We have mentioned
in section 1 that this discrepancy might be due to the threshold setting and sample incompleteness.
We compare the 2-dimensional distribution derived from our simulations (the gray contour plot) with
that predicted by the USJ model (the line contour plot) in Figure 2. The labels marked in Figure 2
indicate the probabilities of a burst within the corresponding regions predicted by the USJ model,
and the filled colors from light gray to dark gray represent the same probability regions derived
from our simulations. From Figure 2, we find that our simulation result is consistent with that
predicted by the USJ model. To clearly illustrate this consistency, we also compare the $1\sigma$
region (the probability of a burst in this region is $\sim 0.68$) of the two distributions in
Figure 3. It is found that the two distributions are most overlap. The results shown in Figures 2
and 3 well agree with that shown in Figure 1.

\begin{figure}
\begin{center}
\includegraphics[width=3.5in,angle=0]{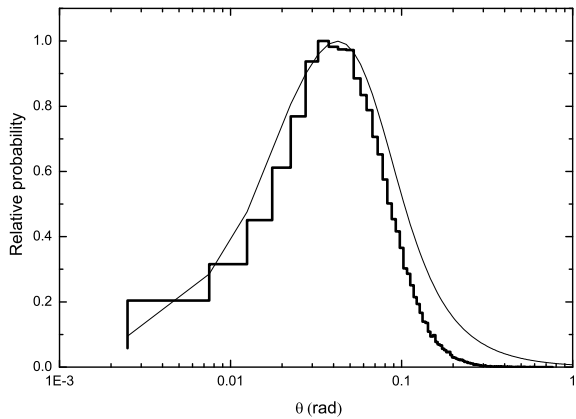}
\end{center}
\caption{The comparison of the relative probability distributions derived from our simulations (the
step line) with that predicted by the USJ model (the straight line).}
\end{figure}

%\begin{figure}
%\begin{center}
%\includegraphics[width=3.5in,angle=0]{fig2.eps}
%\end{center}
%\caption{The $P_{\rm{K-S}}$ as a function of $\theta$.}
%\end{figure}

\begin{figure}
\begin{center}
\includegraphics[width=3.5in,angle=0]{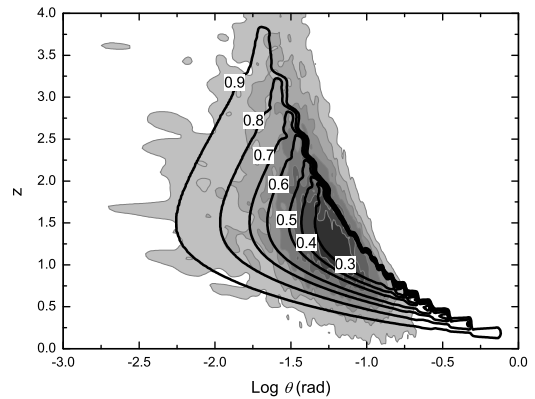}
\end{center}
\caption{The comparison of the 2-dimensional distribution of the relative probability derived from
our simulations (the gray contour) with that predicted by the USJ model (the line contour ).}
\end{figure}

\begin{figure}
\begin{center}
\includegraphics[width=3.5in,angle=0]{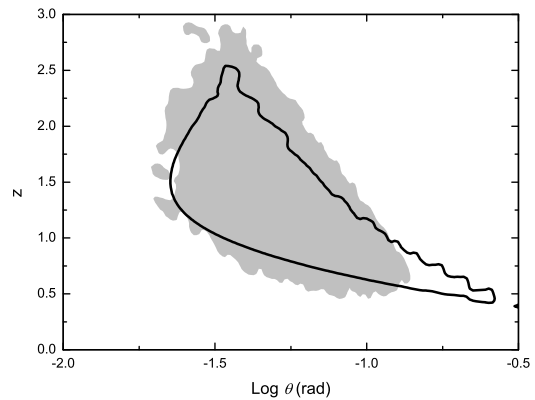}
\end{center}
\caption{The comparison of the $1\sigma$ region of the relative probability distribution derived
from our simulations (the gray contour) with that predicted by the USJ model (the line contour ).}
\end{figure}

\section{ Discussion and Conclusions}
Based on the standard energy reservoir of GRB jets and the relationship between $E_p$ and
$E_{\rm{iso}}$, we test $P(\theta)$ and $P(\theta,z)$ predicted by the USJ model with
observational-based simulations for a detection threshold of $S>4 \times 10^{-5}$ erg cm$^{-2}$. We
simulate a sample including $10^{6}$ GRBs, and then derive $P(\theta)$ and $P(\theta,z)$. We find
that both simulated and theoretical $P(\theta)$ and $P(\theta,z)$ are consistent with each other.

As can be seen in Figures 1-3, a small difference between simulated and theoretical $P(\theta)$ and
$P(\theta,z)$ results from the range of $\theta>0.1$ rad. This difference increases as $\theta$
increases when $\theta>0.1$ rad. The accumulative probability of $\theta<0.25$ rad derived from our
simulations is $\sim 99\%$, while it is 0.83 predicted by the USJ model.  From Eq. (3) one can see
that a burst with lower $E_p$ tends to have a larger $\theta$. Thus, we suspect that the difference
might be due to the limitation of our simulations and the observational biases of $E_p$
distribution. In our simulations, we derive $\theta$ for each burst by Eq. (3), which requires
$E_p(1+z)>100(\varepsilon_{52}/k)^{0.5}\sim 32(\varepsilon_{51}/k)^{0.5}$ keV, where
($\varepsilon_{51}/k)^{0.5}\sim 1$. Those bursts with $E_p(1+z)<32$ keV is excluded from our
simulated GRB sample. However, this limitation could not significantly affect our results since the
$E_p$ distribution used in our simulations mainly ranges in $E_p>100 $ keV. The other possible
reason might be the observational biases of the observed $E_p$ distribution used in our
simulations. For a burst with $\theta>0.25$ rad, its $E_p$ value is less than $100$ keV, assuming
that $\varepsilon_{52}=0.133$, $k=1$, and $z=1$. Thus, the difference between the simulated and
theoretical results at $\theta>0.25$ rad might be the bias of $E_p$ distribution at $E_p<100$ keV.
The distribution is taken from BATSE Large Area Detector (LAD) observations. The BATSE/LAD
thresholds are sensitive to the spectral shape of a burst: the threshold for a hard burst is lower
than the threshold for a softer burst \footnote{http://www.batse.msfc.nasa.gov/batse/}. Bursts with
low $E_p$ seems to be more easily lost by BATSE/LADs. The BATSE/LADs operate in the 20-2000 keV
band but usually trigger in the 50-300 keV band (e.g., Band 2003). BATSE generates a GRB record
when the observed photon fluxes of 2 or 3 energy bands (normally 50-100 keV and 100-300 KeV bands)
are simultaneously beyond the thresholds. In our simulations, we take a threshold as
$F_{th}=S/T=3.3\times 10^{-6}$ erg cm$^{-2}$ s$^{-1}$. We examine whether or not this flux is
beyond the LAD thresholds. We assume that the GRB spectral shape is the Band function (Band 2003)
with a low-energy index $\alpha=-1$ and a high-energy index $\beta=-2$, and then translate the
$F_{th}$ in units of erg cm$^{-2}$ s$^{-1}$ to photon cm$^{-2}$ s$^{-1}$ in a timescale of 1.024
seconds. We obtain 4.57 and 3.29 photon cm$^{-2}$ s$^{-1}$ in the $50-100$ keV and $100-300$ keV
bands when $E_p=100$ keV, respectively. A normal GRB with such a photon flux is intense enough
triggering the BATSE/LADs. However, The observed $E_p$ distribution by BATSE/LADs mainly ranges in
$100-1000$ keV. This might suggest that BATSE/LAD sensitivity decreases significantly for $E_p<100$
keV (Band 2003), and bursts with $E_p<100$ keV are easily lost. If it is really the case, the $E_p$
distribution in our threshold setting might still suffer a little of biases for $E_p<100$ keV. The
upcoming {\em Swift} satellite, which will be scheduled for launch in 2004 September (Gehrels
2004), is marginally more sensitive than BATSE for $E_p>100$ keV but significantly more sensitive
for $E_p<100$ keV (Band 2003). It might establish a more reliable $E_p$ distribution, which may be
used for a further examination of our results.

The true GRB rate as a function of redshift is difficult to determine from the current GRB sample.
We assume that the GRB rate is proportional to the star formation rate. Various models of star
formation rates are presented in the literature, and they are quite different. In our analysis, we
focus on the comparison of the simulated $\theta$ distribution with that predicted by the USJ
model. We find that the comparison result is not significantly affected by the redshift
distribution. In our analysis, the SF1 model from Porciani \& Madau (2001) is used. We use the SF2
model (Porciani \& Madau 2001) to perform the simulations. The results are shown in Figure 4. One
can see that the comparison of the two distributions is not significantly affected by the model of
redshift distribution.

\begin{figure}
\begin{center}
\includegraphics[width=3.5in,angle=0]{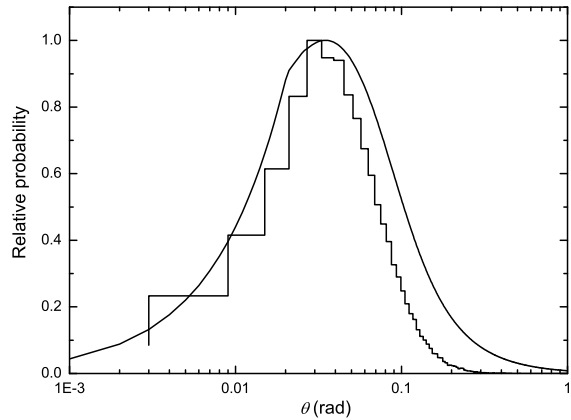}
\end{center}
\caption{Same as Figure 1 but SF2 model is used.}
\end{figure}

Based on our results and the above discussion, we conclude that both the probability distributions,
$P(\theta)$ and $P(\theta,z)$, derived from our simulations are consistent with that predicted by
the USJ model. This consistency may be regarded as a support of USJ model.

We would like to address our great thanks to the anonymous referee for his thoughtful comments and
helpful suggestions, which have allowed us to improve this revised paper greatly. We also thank
Bing Zhang and Xinyu Dai for their helpful discussion. This work was supported by the National
Natural Science Foundation of China (grants 10233010 and 10221001), the National 973 Project
(NKBRSF G19990754), the Natural Science Foundation of Yunnan (2001A0025Q), and the Research
Foundation of Guangxi University.

\end{document}